# An alternative approach to Schrödinger equations with a spatially varying mass

M. Çapak and B. Gönül

Department of Engineering Physics, Faculty of Engineering, University of Gaziantep, 27310, Gaziantep-Turkey

**Abstract**

Extending the point canonical transformation approach in a manner distinct from the previous ones, we propose a unified approach of generating potentials of all classes having non-constant masses.

## 1. INTRODUCTION

Tracking down solvable quantum potentials has always aroused interest. Apart from being useful in understanding of many physical phenomena, investigatations of such potentials also provide a good starting point for undertaking perturbative calculations of more complex systems. Within this context, we have recently suggested [1] a novel algebraic framework for the unified treatment of Schrödinger equations, with a constanst mass, involving solvable and non-solvable potentials.

For the completeness, and also considering the importance of quantum mechanical sytems with position dependent masses in describing the physics of many microstructures of current interest [2], the model discussed in Ref. [1] is to be extended in this letter for also analysing the sytems having non-constant masses.

Unfortunately, up to now, the related previous works have dealt with only either exact or quasi-exact solutions of the Schrödinger equation in a position-dependent effective mass back-ground. In the other words, to the best of our knowlege, there has been no ongoing discussion in the literature regarding approximate solutions of non-solvable potentials for the case of non-consant masses. Therefore, at present, there seems no room for a precise test of the application results of the prescription introduced in Section 2. However, we believe that the present algebraic scenario would find a widespread application in the near future due to the current interest in microstructures, Ref [2], which eventually would require physically acceptable approximate descriptions of such systems that should be investigated within the frame of more complex and non-exactly solvable potentials.

In addition, the recent progress in analysing experimental data within the frame of algebraic models to investigate nuclear structure gives us a signal for the possibility of using a prescription such as the one presented in this article [3,4].

## 2. FORMALISM

As is well known the general Hermitian position-dependent effective mass Hamiltonian, initially proposed by von Roos [5] in terms of ambiguity parameters $\alpha$, $\beta$, $\gamma$ such that $\alpha + \beta + \gamma = -1$, gives rise to the time independent Schrödinger equation

$$H\psi(x) = \left[-\frac{d}{dx}\frac{1}{M(x)}\frac{d}{dx} + V_{eff}(x)\right]\psi(x) = E\psi(x) \quad, \tag{1}$$

where the effective potential

$$V_{eff}(x) = V(x) + \frac{1}{2}(\beta+1)\frac{M''}{M^2} - [\alpha(\alpha+\beta+1) + \beta + 1]\frac{M'^2}{M^3} \tag{2}$$

depends on the mass term. Here the prime denotes derivative with respect to $x$, $M(x)$ is the dimensionless form of the mass function $m(x) = m_0 M(x)$ and we have set $\hbar = 2m_0 = 1$.

Considering the suggestion in Ref. [1]

$$\psi(x) = [f(x)F(g(x))]h(x) \quad, \tag{3}$$

for the general solution of second order differential equations such as Eq. (1), where $f(x)F(g)$ yields an algebraic closed solution for exactly and quasi-exactly solvable potentials with $F(g)$ being a special function satisfying Schrödinger-like equations

$$\frac{d^2F}{dg^2} + Q(g)\frac{dF}{dg} + R(g)F(g) = 0 \quad, \tag{4}$$

while $h(x)$ is the moderating function in connection with a perturbing piece of the full potential corresponding to, in our present consideration, Eq. (2). The form of $Q(g)$ and $R(g)$ is already well defined for any special function $F(g)$ when dealing with analytically solvable potentials. However, in case of the consideration of a realistic non-exactly solvable system one should suggest reliable expressions, in an explicit form, for plausible definitions of the related $Q(g)$ and $R(g)$ functions. This is the significant point in the framework of the new formalism to reach physically meaningful solutions.

On inserting Eq. (3) in Eq. (1) and comparing the result with Eq. (4), we arrive at

$$Q(g(x)) = \frac{g''}{(g')^2} + \frac{2f'}{fg'} - \frac{M'}{Mg'} + \frac{2h'}{hg'} \qquad (5)$$

and

$$R(g(x)) = \frac{f''}{f(g')^2} + \frac{2f'h'}{fh(g')^2} + \frac{h''}{h(g')^2} - \frac{M'}{M}\frac{f'}{f(g')^2} - \frac{M'}{M}\frac{h'}{h(g')^2} + M\frac{[E - V_{eff}(x)]}{(g')^2} \qquad . \qquad (6)$$

where

$$f(x) \propto \left(\frac{M}{g'}\right)^{1/2} \exp\left[\frac{1}{2}\int^{g(x)} Q(y)dy\right] \qquad . \qquad (7)$$

Obviously, Eqs. (5-7) reduce to their standard forms (see, e.g., Ref. [6]) for the consideration of exact solvability of Schrödinger equations with a constant mass in which case $M(x)$ and $h(x)$ in the equations above appear as a constant. Gaining confidence from this observation we proceed with

$$V_{eff}(x) = V_{ES}(x) + \Delta V(x)$$
$$E = E_{ES} + \Delta E \qquad (8)$$

in accordance with our choice in (2), which means that potentials considered in this article are admitted as the sum of an exactly (or quasi-exactly) solvable potential with a perturbation or a moderating piece. Hence, the aim in this perspective is to reveal the corrections to energy ($\Delta E$) and wavefunction, $h(x)$, for a given $\Delta V(x)$ as the main piece of the solutions leading to exact solvability can easily be found from the literature.

The use of Eqs. (5) and (8) within the frame of Eq. (6) leads to the coupled equations in the form of

$$E_{ES} - V_{ES}(x) = \frac{g'''}{2Mg'} - \frac{3}{4M}\left(\frac{g''}{g'}\right)^2 + \frac{(g')^2}{M}\left[R_{ES}(g) - \frac{1}{2}\frac{dQ_{ES}(g)}{dg} - \frac{1}{4}Q_{ES}^2(g)\right] - \frac{M''}{2M^2} + \frac{3M'^2}{4M^3}$$

(9)

and

$$\Delta E - \Delta V(x) = -\frac{1}{2M}\left(g'' + \frac{2f'g'}{f} - \frac{M'g'}{M}\right)\Delta Q(g) + \frac{g'^2}{M}\left[\Delta R(g) - \frac{1}{2}\frac{d(\Delta Q)}{dg} - \frac{1}{4}\Delta Q^2(g)\right] \quad (10)$$

in which, from Eq. (5), $\Delta Q(g) = \dfrac{2h'}{hg'}$ leading to

$$h(x) = \exp\left(\frac{1}{2}\int \Delta Q(g)\,dg\right), \quad (11)$$

and $\Delta R(g) = \dfrac{2f'h'}{f\,h(g')^2} + \dfrac{h''}{h(g')^2} - \dfrac{M'}{M}\dfrac{h'}{h(g')^2} + M\dfrac{[\Delta E - \Delta V(x)]}{(g')^2}$. It is clear that $Q_{ES}(g)$, together with $R_{ES}(g)$, related to algebraically solvable potentials, can be obtained from (5) and (6) such that $Q(g) = Q_{ES}(g) + \Delta Q(g)$ and $R(g) = R_{ES}(g) + \Delta R(g)$. Again, in case of exact solvability Eqs. (10) and (11) disappear naturally by reducing the scheme to Eq. (9), which justifies the reliability of the present work, see, e.g., Ref. [7]. This realization puts forward the significance of Eq. (10) for approximately solvable more complex quantum potentials, which is the main point in this letter.

The result of this brief investigation opens a gate to the reader for the visualization of the explicit form of the correction ($\Delta E$) to the energy. Unfortunately, there seems a problem arised in calculating the correction term owing to the presence of two unknown: $\Delta Q(g)$ and $\Delta R(g)$ on the right hand side of Eq. (10). To circumvent the resulting drawback and proceed safely, one needs to use the interesting expression yielding inter-connection between $\Delta Q$ and $\Delta R$ functions

$$\Delta R(g) = -\Delta Q(g)\frac{F'(g)}{F(g)}, \quad (12)$$

that is obtained, after some exhaustive analyses, by considering the another form of Eq. (10)

$$\Delta E - \Delta V(x) = -\frac{1}{M}\left[\frac{h''}{h} + \frac{2h'}{h}\left(\frac{f'}{f} + \frac{g'F'(g)}{F(g)} - \frac{M'}{2M}\right)\right]. \quad (13)$$

As $F(g)$ is well defined for solvable potentials, we obviously need here to find only an appropriate expression for $\Delta Q(g)$ to be employed eventually in (10), or in (13) with the use of (11), that reveals the full solution.

It is however reminded that if the whole potential such as in (2) has no analytical solution, one should expand the related functions in terms of the perturbation such that $\Delta V(x) = \sum_{N=1}^{\infty}\Delta V_N(x)$ and $\Delta E_n = \sum_{N=1}^{\infty} \Delta E_{nN}$ where $N$ denotes the perturbation order and $n(=0,1,2,...)$ is the radial quantum number relating to bound-state energy levels. In connection with this idea, the form of $\Delta Q$ should be chosen carefully depending on $N-$values and the substitution of reasonable $\Delta Q$ values into either (10) or (13), and consequently equating terms with the same power of the perturbation order on both sides, yields the modifying terms in the frame of coupled equations at successive orders for different states. The procedure has been well discussed in Ref. [1] for approximately solvable unharmonic potentials involving constant masses.

## 3. APPLICATION

This section gives a clue for exploring new potentials within the unique frame of the formalism introduced. For clarity, we start with a simple example from the work of Bagchi and his co-workers [7] where the form of the mass is chosen as $M = \lambda g'$ with $\lambda$ being a constant, which simplifies Eq. (9) as

$$E_{ES} - V_{ES}(x) = \frac{g'}{\lambda}\left[R_{ES}(g) - \frac{1}{2}\frac{dQ_{ES}(g)}{dg} - \frac{1}{4}Q_{ES}^2(g)\right], \tag{14}$$

that is responsible for the piece of the interaction having closed analytical solutions. Another appropriate choice for the $F(g)$ within the list of the orthogonal polynomials used frequently in physics, $F(g) = \exp\left(-\frac{g}{2}\right)g^{\left(\frac{\alpha+1}{2}\right)}L_n^\alpha(g)$, implies that

$$R(g) = \frac{2n+\alpha+1}{2g} + \frac{1-\alpha^2}{4g^2} - \frac{1}{4} \quad , \quad Q(g) = 0 \quad . \tag{15}$$

The use of (15) in (14) requires that $g'/\lambda g = \beta^2$ which must be restricted to positive values in order to get a constant term on the right-hand side of Eq. (14). Setting $\lambda = -1/\beta$, we arrive at $g(x) = e^{-\beta x}$ and, from the definition of the mass form, $M(x) = e^{-\beta x}$. Therefore, the related solutions appear as

$$E_{ES} = \beta^2\left(n + \frac{\alpha+1}{2}\right) \quad , \quad \psi_{ES} = f(x)F(g) \propto \exp\left(-\frac{g}{2}\right)g^{\left(\frac{\alpha+1}{2}\right)}L_n^\alpha(g) \quad , \tag{16}$$

for the constructed solvable potential

$$V_{ES}(x) = \frac{\beta^2}{4}\left[(\alpha^2-1)e^{\beta x} + e^{-\beta x}\right] \quad . \tag{17}$$

At this stage, we focus on Eq. (10) to observe how additional potential terms can be generated. This observation would clarify how the present model copes with the perturbed piece of the potential, as well as the understanding of the procedure for the generation of new non-solvable potentials. With the use of $M = \lambda g'$ in (10), we obtain

$$\Delta E - \Delta V(x) = \frac{g'}{\lambda}\left[\Delta R(g) - \frac{1}{2}\frac{d(\Delta Q)}{dg} - \frac{1}{4}\Delta Q^2(g)\right], \tag{18}$$

which is similar to Eq. (14). Bearing the structure of (14) in mind, in particular $n$-dependence of $R(g)$ term, we transform the above equation, with the remind of Eq. (12),

$$\Delta E - \Delta V(x) = -\frac{g'}{\lambda}\left[\frac{F'}{F}\Delta Q + \frac{1}{2}\frac{d(\Delta Q)}{dg} + \frac{1}{4}\Delta Q^2\right], \tag{19}$$

into the more applicable form. As the study of (14) has already defined $F(g)$, $g(x)$ and $\lambda$, one needs here to deal with only $\Delta Q(g)$ term which is the central part of the whole discussion. Obviosly different choice of this term, depending of course on the perturbation order ($N$) discussed in the previos section, would lead to distinct functions generating new additional potentials including the corresponding energy terms in each order. For instance, a possible choice of $\Delta Q(g) = -\frac{b}{g'}\sum_{i=1}^{N}G(i,x)$, where $b$ is a constant and $G$ is a general form of the physically reasonable functions, would help us to define energy term from the first term in the bracket while generating modifications to the main part of the potential in (17). It is clear that there is no strict definition for the generating function $G(i,x)$ which should be determined by considering the structure of the realistic system of interest. It is however stressed that the modifications in each order for successive quantum levels should be indivudually considered. This requires a meticulous search for the corrections brought by (19), in which expressions giving the structure of $F(g)$ for each state $(n = 0, 1, 2, ...)$ ought to be used properly.

Finally, to clarify the flexibility of the approach used, it should be noticed that additional exactly and quasi-exactly solvable quantum potentials in a spatially varying mass context may also be generated. As an illustration, we choose again $M = \lambda g'$ option and employ it in Eq. (10). This consideration leads to

$$\Delta E - \Delta V(x) = -\frac{\Delta Q(g)}{\lambda}\frac{[\psi_{ES}(x)]'}{\psi_{ES}(x)} - \frac{g'}{2\lambda}\frac{d[\Delta Q(g)]}{dg} - \frac{g'}{4\lambda}[\Delta Q(g)]^2, \tag{20}$$

where, from Eq. (3), $\psi_{ES}(x) = f(x)F(g)$ is the exact (or quasi-exact) solution for a solvable potential undertaken. To proceed we choose a specific example assuming that the sum of the second and third terms on the right hand side of the above equation is zero. This feasible assumption requires that $\Delta Q = 2/g$ which transforms Eq. (20) to

$$\Delta E - \Delta V(x) = -\frac{2}{\lambda g} \frac{[\psi_{ES}(x)]'}{\psi_{ES}(x)} . \qquad (21)$$

As, in general, $\psi_{ES}$ and $g(x)$ are readily obtained from Eq. (9), or they can be extracted from the literature, Eq. (21) together with Eq. (8) may generate new solvabe potentials. In addition, the modificiation to the wavefunction $(h(x))$ due to the additional potential $(\Delta V(x))$ can easily be obtained through Eq. (11). More specifically, the substitution of $\psi_{ES}$ in (16) into Eq. (21) generates a modification in a closed form to the solvable potential in (17), the summation of these terms $(V_{ES} + \Delta V)$ is a new potential having algebraic solutions involving either the whole or some part of the spectra.

Overall, the method described here has an ability of generating all classes of potentials depending of course on the choice of the mass function and the use of different orthogonal polynomials $(F(g))$ within the frame of Eqs. (9) and (12), providing closed analytical or approximate solutions to the corresponding Schrödinger equation.

## 4. CONCLUDING REMARKS

The literature covers many applications regarding the description of collective nuclear properties in terms of the corresponding collective variables within the framework of Bohr Hamiltonian involving exactly solvable potentials. However, the recent analysis [3] has introduced special solutions for the Morse potential which is known to be exactly soluble only for $\ell = 0$, unlike the previous applications. For the approximate treatment of quantum states having non-zero angular momenta, the well-known Pekeris approximation has been used. This unusual consideration, however, has suppressed the overestimation of the energy spacings within the beta-band due to right branch of the Morse potential that imitates the sloped wall, which has removed a main drawback of the earlier considerations. Furthermore, the Bohr Hamiltonian and its extensions, for a recent review see Ref. [3] and the references therein, have provided for several decades a sound framework for understanding the collective behaviour of atomic nuclei. It has been customary to consider in the Bohr Hamiltonian the mass to be constant, as in [3]. However, evidence has been accumulating that this approximation might be inadequate [4] in which significant effects of using a mass depending upon nuclear deformation on the calculations of spectra and analytical expression of the wavefunction decribing the collective motion of deformed nuclei was discussed in detail. Within this context, the present algebraic model seems promising as the scheme has a power of yielding the required expressions in an explicit form for the Morse-like potentials mentioned above and also the corrections due to the corresponding angular momentum barrier involving deformation-/position-dependent mass systems. Along this line the works are in progress.

Moreover, recently, a modified factorization technique [8] of a quantum system characterized by spatially varying mass-Hamiltonians has been presented and shown that excited state wavefunctions of a given singular Hamiltonian can be used to construct non-singular isospectral partner potentials. This work would be helpful to remove the singularity problem naturaly arised in the present formalism, Eqs (10), (12) and (13), due to zeros of $F(g)$ function, which should be necessary for the case of considering higher quantum states.